\documentclass[english,12pt,preprint]{revtex4}
\usepackage[T1]{fontenc}
\usepackage[latin1]{inputenc}
\usepackage{graphicx}

\makeatletter

\usepackage{babel}
\makeatother
\begin{document}

\title{Controlling a Telescope Chopping Secondary Mirror Assembly Using
a Signal Deconvolution Technique}

\author{Martin Houde, Lynn C. Holt, Hiroshige Yoshida, and Patrick M. Nelson}

\affiliation{Caltech Submillimeter Observatory, 111 Nowelo Street, Hilo, HI 96720}

\begin{abstract}
We describe a technique for improving the response of a telescope
chopping secondary mirror assembly by using a signal processing method
based on the Lucy deconvolution technique. This technique is general
and could be used for any systems, linear or non-linear, where the
transfer function(s) can be measured with sufficient precision. We
demonstrate how the method was implemented and show results obtained
at the Caltech Submillimeter Observatory using different chop throw
amplitudes and frequencies. No intervention from the telescope user
is needed besides the selection of the chop throw amplitude and frequency.
All the calculations are done automatically once the appropriate command
is issued from the user interface of the observatory's main computer.
\end{abstract}
\maketitle

\section{Introduction}

Chopping scans are widely used in radioastronomy as they provide an
efficient way to reduce the adverse effects that any instabilities
present either in the sky signal or some telescope equipment can have
on the detection of weak signals. A chopping scan is defined as a
mode of observation where the telescope's secondary mirror is rotated
back and forth through some angle and where the signals from both
{}``end'' positions are integrated separately and subtracted from
each other. This mode is to be compared with the so-called ON-OFF
position (beam switching) scan where the telescope actually moves
back and forth from one end position to the other. Because of the
much greater speed at which the secondary mirror can move compared
to the telescope, the signal subtraction happens much faster and thus
an increase in the ability to detect weak signals. By moving or chopping
the secondary mirror even at a relatively low frequency (e.g., 1 Hz)
one can obtain a significant improvement in baseline quality when
compared to a typical beam switch. In what will follow, the secondary
mirror displacements are in units of arcseconds as measured on the
sky.

At the Caltech Submillimeter Observatory (CSO) a chopping secondary
mirror assembly was installed in 1994 and has since been used both
for heterodyne receivers and bolometer cameras (e.g., SHARC and HERTZ)
observations. It is composed, in part, of a carbon fiber mirror mounted
on a DC brushless motor along with a system of counterweights which
greatly reduces the amount of vibration noise transmitted to the observatory's
receivers or cameras. The huge advantage that this vibration suppression
technique brings, for the detection of weak signals comes, however,
at the cost of an increase in the inertia of the chopper assembly
which causes a reduction in the speed and an increase in the settling
time in the response of the system.

We show in Figure \ref{cap:pid} a block diagram of the chopping secondary
system as it has been used since its installation at the CSO. Once
the user of the telescope has selected a chop throw amplitude (in
arcseconds) and frequency, a square wave is sent to the input of a
typical Proportional-Integral-Derivative (PID) electronic controller
\citep{Ogata} where it is compared to the position signal of the
mirror (obtained through a Linear Variable Differential Transformer
or LVDT). The processed error signal is then sent to a power amplifier
which feeds the motor and thus continuously re-positions the mirror
while the PID controller acts to minimize the error signal.

Because of the relatively slow response of the chopper assembly, and
the non-linearities inherent to the system (see section \ref{sec:Implementation}),
the parameters of the PID controller cannot be held fixed at a given
set of values but have to be adjusted by the user of the telescope
for different chop throw amplitudes and frequencies. Although this
does not present a problem in principle it has been the experience
that the tuning of the controller's parameters can sometimes be a
time consuming effort that reduces the efficiency of the observatory.
Also, since, as will be shown later, the response time of the assembly
is of the order of the chopping period (or more) it is often quite
difficult to find the appropriate set of parameters that will give
optimum results. Too often, the outcome of such exercise is a reduction
in the performance of the chopping assembly; both in its settling
time and positioning accuracy.

In the following sections of this paper we will demonstrate how a
signal processing method based on the Lucy deconvolution technique
\citep{Lucy} was implemented at the CSO to solve this problem and
provide a system that requires no intervention from the telescope
user, while keeping hardware changes to a minimum. We will start in
the next section with a brief exposition of the set of equations that
govern the Lucy deconvolution technique followed by a presentation
of the new chopping secondary assembly (section \ref{sec:new-assembly}).
We will finish by showing how the deconvolution technique was implemented,
along with the needed modifications, and by presenting some results
obtained so far.

\section{Lucy's Deconvolution Technique}

An iterative method for signal deconvolution based on the Bayes rule
for conditional probabilities was introduced by \citet{Lucy} and
has been successfully used in astronomy for the processing and extraction
of precise photometric information from originally blurred images
taken under average seeing conditions (see for example \citet{Houde}).

Limiting ourselves to a one-dimensional problem, the set of equations
governing Lucy's technique is relatively simple. Denoting by $r\left(t\right)$
and $s\left(t\right)$ the input and output signals of a linear system,
respectively, we know that they are related to each other through
the transfer function $h\left(t\right)$ of this same system by the
following convolution integral

\begin{equation}
s\left(t\right)=\int r\left(\tau\right)h\left(t-\tau\right)d\tau,\label{eq:conv}\end{equation}

\noindent where the limits of integration in equation (\ref{eq:conv}),
and in all of the integrals that will follow, are from $-\infty$
to $+\infty$. 

The goal of a deconvolution technique is to invert equation (\ref{eq:conv})
and express $r\left(\tau\right)$ as a function of $s\left(t\right)$
using a new function $g\left(\tau-t\right)$ as follows

\begin{equation}
r\left(\tau\right)=\int s\left(t\right)g\left(\tau-t\right)dt.\label{eq:deconv}\end{equation}

Lucy's idea was to liken the (reversed) time shifted transfer function
$h\left(t-\tau\right)$ to a Bayes density function of conditional
probability. In doing so, the new function $g\left(\tau-t\right)$
can be interpreted as a new density function and readily determined
using the Bayes rule by \citep{Haykin}

\begin{equation}
g\left(\tau-t\right)=\frac{r\left(\tau\right)h\left(t-\tau\right)}{s\left(t\right)},\label{eq:g}\end{equation}

\noindent or alternatively

\[
g\left(\tau-t\right)=\frac{r\left(\tau\right)h\left(t-\tau\right)}{\int r\left(\lambda\right)h\left(t-\lambda\right)d\lambda}.\]

Evidently, it is impossible to directly determine $g\left(\tau-t\right)$
with equation (\ref{eq:g}) since it is expressed as a function of
$r\left(\tau\right)$ which is the unknown that we are trying to evaluate.
But the form of equations (\ref{eq:conv}), (\ref{eq:deconv}) and
(\ref{eq:g}) suggests a simple iterative method that can be used
to solve the problem.

If we supply an initial {}``guess'' $r_{0}\left(\tau\right)$ for
$r\left(\tau\right)$ and insert it in equation (\ref{eq:conv}),
we find a first approximate solution $s_{0}\left(t\right)$ to $s\left(t\right)$.
We then in turn insert $s_{0}\left(t\right)$ along with $r_{0}\left(\tau\right)$
in equation (\ref{eq:g}) to get an approximation $g_{0}\left(\tau-t\right)$
for $g\left(\tau-t\right)$. Finally, $g_{0}\left(\tau-t\right)$
is used in equation (\ref{eq:deconv}) to get a new function $r_{1}\left(\tau\right)$,
and so on. This process can be repeated as often as desired or until
convergence is attained.

The final set of equations that define this iterative algorithm can
then be written as follows

\begin{eqnarray}
s_{i}\left(t\right) & = & \int r_{i}\left(\tau\right)h\left(t-\tau\right)d\tau\label{eq:s_i}\\
r_{i+1}\left(\tau\right) & = & r_{i}\left(\tau\right)\int\frac{s\left(t\right)}{s_{i}\left(t\right)}h\left(t-\tau\right)dt.\label{eq:r_i+1}\end{eqnarray}

\noindent for $i=0,1,2,...$ 

Finally, two comments to end this section:

\begin{itemize}
\item the applicability of the solution to the problem given by equations
(\ref{eq:s_i}) and (\ref{eq:r_i+1}) is based on the implied assumption
that the transfer function of the system $h\left(t\right)$ can be
measured independently or is known \emph{a priori}. This is true for
the problem of the chopping secondary that will be addressed starting
in the next section
\item it will be noted that the integral in equation (\ref{eq:r_i+1}) is
actually a correlation. It follows that the algorithm dictated by
the final set of equations can easily be programmed (i.e., computer
coded) using subroutines based on the so-called Fast-Fourier-Transform
(FFT) methods for convolution and correlation integrals. This is what
we have done in the implementation of our technique where we have
used Fortran routines presented by \citet{NR}.
\end{itemize}

\section{The new CSO Chopping Secondary Mirror Assembly\label{sec:new-assembly}}

In a simple implementation of equations (\ref{eq:s_i}) and (\ref{eq:r_i+1})
one needs a way to generate an input signal to be applied to a given
system, measure the output of the system when subjected to this input
and finally evaluate the transfer function of the system. In order
to accomplish this with our chopping secondary system we modified
our assembly from that of Figure \ref{cap:pid} to that of Figure
\ref{cap:new_pid}. We have replaced the square wave generator of
our original system by a Real Time Linux (RT Linux) computer which
is equipped with the necessary input/output devices (i.e., Analog-to-Digital
and Digital-to-Analog converters) to achieve these tasks.

The RT Linux computer also serves as host to the program that performs
the necessary calculations and measurements that will allow for the
determination of the optimum input to the chopper assembly.

Ideally the sequence of operations would go like this

\begin{enumerate}
\item Calibration of the system: signals of constant level are sent to the
input of the assembly and the corresponding output levels are measured.
In this manner, the {}``gain'' and {}``offset'' of the system
are determined and applied to all subsequent input/output operations.
\item Evaluation of the transfer function: this is done by sending a step
input signal of a given amplitude to the chopper assembly and calculating
the normalized time derivative of the corresponding output signal.
This is a very simple way to evaluate a transfer function since the
convolution of an arbitrary function with a unit step function is
equivalent to the primitive of the original function. This is the
technique we use although it should be noted that we also smooth the
resulting time derivative with a Savitzky-Golay filter \citep{NR}
to reduce the impact of noise in the application of the Lucy deconvolution
technique. We will show some examples of measured transfer functions
in the next section.
\item Determination of the desired or targeted output signal $s\left(t\right)$.
\item Determination of the optimum input signal: to do so one would \emph{i)}
choose an arbitrary waveform as a hypothetical input of the assembly
($r_{0}\left(\tau\right)$ in equations (\ref{eq:s_i}) and (\ref{eq:r_i+1})),
\emph{ii)} calculate the corresponding output response $s_{0}\left(t\right)$
of the system using equation (\ref{eq:s_i}) and \emph{iii)} calculate
a new input $r_{1}\left(\tau\right)$ using equation (\ref{eq:r_i+1}).
Repeat \emph{ii)} and \emph{iii)} (using $r_{i}\left(\tau\right)$
and $s_{i}\left(t\right)$, with $i=1,2,...$, instead of $r_{0}\left(\tau\right)$
and $s_{0}\left(t\right)$) until convergence to the best input $r_{f}\left(\tau\right)$
signal is attained.
\item Finally, $r_{f}\left(\tau\right)$ is applied to the input of the
assembly to produce the output $s_{f}\left(t\right)$ that most resemble
the desired output $s\left(t\right)$.
\end{enumerate}
We have tested this technique on simple linear systems (e.g., electrical
RC filters) with very good results. However, when applied to our chopping
secondary mirror assembly the technique did not work in general. It
was determined that the non-linearities in the system's response were
the cause of this failure and forced us to bring some changes to the
algorithm discussed here.

\section{Implementation of the Method\label{sec:Implementation}}

\subsection{Non-linearities}

Since we have a DC motor as one of the main components of the chopper
assembly, it is not surprising that the system should include some
non-linearities in its response. As one should expect, the magnetic
core of the motor is inherently non-linear as it will experience different
amounts of saturation depending on the amplitude of the excitation
it is subjected to. That is to say that the transfer function of the
system changes with the input signal and that the system reacts differently
to different chop throw amplitudes. Moreover, it is also the case
that the sign of chop throw will affect the shape of the transfer
function. Simply stated, the system has hysteresis and, therefore,
does not go {}``up'' the same way it goes {}``down''.

This will be made clear with the results presented in Figure \ref{cap:tfunc}.
In this figure we can see the effect that the non-linearities have
on the transfer function of the system. The transfer functions shown
were measured using the method discussed earlier using rising ({}``up'')
and falling ({}``down'') step functions (90 arcseconds in amplitude)
at two different rest positions (0 and 180 arcseconds for the top
and bottom graphs, respectively).

From this it is clear why the algorithm defined by equations (\ref{eq:s_i})
and (\ref{eq:r_i+1}) would be doomed to failure. The same two equations
can, however, be easily adapted to the problem at hand and make it
possible to use the Lucy deconvolution method (albeit somewhat modified)
for this kind of non-linear system.

\subsection{Modifications to the Lucy Deconvolution Method}

As was mentioned in the last section, the fact there does not exist
a single transfer function that defines the system does not imply
that we cannot use the Lucy deconvolution technique to achieve our
goal, but we must acknowledge the existence of a family of transfer
functions that are dependent on the input signal to the system. That
is to say, we should replace $h\left(t\right)$ by $h\left(t;r\right)$,
the aforementioned dependence on the input signal $r\left(t\right)$
now being made explicit. In practice this means that we now have to
measure the transfer functions of the system along a sufficiently
refined two-dimensional grid of different step amplitudes (positive
and negative) and rest positions. Four examples of such measurements
were shown in Figure \ref{cap:tfunc}. For the results that will be
presented later in this section, we have used a grid where the step
amplitude ranges from $-240$ to 240 arcseconds with a resolution
of 30 arcseconds and the rest position spans a similar domain with
half the resolution (i.e., 60 arcseconds). It should be noted that
this measurement of the transfer functions requires a fair amount
of time (as much as 15 to 20 minutes for the grid defined above).
But, on the other hand, it needs to be done only once and does not
have to be repeated for different chop throw amplitudes and frequencies.

Another important thing to realize is that, contrary to instances
where one uses the Lucy technique to deconvolve astronomical images
\citep{Houde}, we are here free to use the system to measure its
response to any given input signal and not forced to calculate it
through equation (\ref{eq:s_i}). This means that the original set
of equations (\ref{eq:s_i}) and (\ref{eq:r_i+1}) can be reduced
to only one equation, namely

\begin{equation}
r_{i+1}\left(\tau\right)=r_{i}\left(\tau\right)\int\frac{s\left(t\right)}{s_{i}\left(t\right)}h\left(t-\tau;r_{i}\right)dt.\label{eq:modLucy}\end{equation}

With these modifications, the sequence of operations defined in section
\ref{sec:new-assembly} now becomes

\begin{enumerate}
\item Calibration of the system: signals of constant level are sent to the
input of the assembly and the corresponding output levels are measured.
In this manner, the {}``gain'' and {}``offset'' of the system
are determined and applied to all subsequent input/output operations.
\item Evaluation of the transfer functions: a set of step input signals
of differing amplitudes and rest positions are sequentially sent to
the chopper assembly and the transfer functions are measured by calculating
the normalized time derivative of the corresponding output signals.
A Savitzky-Golay filter \citep{NR} is applied to the functions to
reduce the impact of noise on the deconvolution.
\item Determination of the desired or targeted output signal $s\left(t\right)$.
\item Determination of the optimum input signal: to do so one would \emph{i)}
send an arbitrary wave form $r_{0}\left(\tau\right)$ to the input
to the assembly, \emph{ii)} measure the corresponding output response
$s_{0}\left(t\right)$ of the system and \emph{iii)} use equation
(\ref{eq:modLucy}) to determine a new input signal $r_{1}\left(\tau\right)$.
Repeat \emph{i)}, \emph{ii)} and \emph{iii)} (using $r_{i}\left(\tau\right)$
and $s_{i}\left(t\right)$, with $i=1,2,...$, instead of $r_{0}\left(\tau\right)$
and $s_{0}\left(t\right)$) until convergence to the best input $r_{f}\left(\tau\right)$
signal is attained.
\item Finally, $r_{f}\left(\tau\right)$ is applied to the input of the
assembly to produce the output $s_{f}\left(t\right)$ that most resemble
the desired output $s\left(t\right)$.
\end{enumerate}
We have applied this technique to our chopping secondary mirror assembly
at the CSO with success. We show typical results in Figures \ref{cap:60_1hz}
and \ref{cap:300_1hz} for chop throws of 60 arcseconds and 300 arcseconds,
respectively, at a frequency of 1 Hertz. For this, we chose the initial
input signal $r_{0}\left(\tau\right)$ to be a square wave with corresponding
amplitudes and frequency, the system's response to this input is labeled
{}``uncorrected output'' in the legends. The desired or {}``targeted
output'' signals corresponding to $s\left(t\right)$ (also shown
on the graphs) in equation (\ref{eq:modLucy}) in both cases rise
(or fall) at the same rate of 3 arcseconds per millisecond when not
constant. A comparison of the {}``uncorrected output'' ($s_{0}\left(t\right)$)
with the {}``corrected output'' ($s_{f}\left(t\right)$) shows the
power of this deconvolution method when applied to this type of problems.
In both cases the improvement is significant. Furthermore, it would
have been next to impossible to guess which form should the final
input signal $r_{f}\left(\tau\right)$ (shown by the {}``applied
input'' curves in the graphs) take to obtained the desired outcome.
The residual error signal is also plotted in the bottom graph of each
figure. As can be seen, the RMS error calculated on flatter parts
of the curves are in both cases small ($\leq1.1$ arcsecond).

Referring to Figure \ref{cap:tfunc} we see that the transfer function
of the system settles down in about 0.5 seconds, which is exactly
equal to half of the period of the signals displayed in Figures \ref{cap:60_1hz}
and \ref{cap:300_1hz}. This means that the assembly would have just
enough time to settle into steady state during half of a cycle when
subjected to a square wave of a frequency of 1 Hertz. It would be
interesting to see how our technique fares when the period of the
input signal is reduced to a value that is significantly less than
that the system's settling time. To test this we have subjected the
chopper assembly to a signal of a frequency of 4 Hz and tried to obtained
an output of 60 arcseconds in amplitude. This is shown on Figure \ref{cap:60_4hz}
where now the {}``targeted output'' rises and falls at a rate of
6 arcseconds per millisecond when not constant. Although as could
be expected the overall shape of the resulting output signal is somewhat
more {}``rounded'' when compared to the results shown in Figure
\ref{cap:60_1hz}, the improvement obtained in going from the {}``uncorrected
output'' to the final output signal (i.e., {}``corrected output''
on the graph) is rather significant. In fact, we can see from the
bottom graph that for about 52\% of a period the response is at most
within a few arcseconds from the desired position; the RMS value of
the error on that portion of the signal is 1.4 arcseconds. 

\begin{acknowledgments}
The Caltech Submillimeter Observatory is funded by the NSF through
contract AST 9980846. The chopping secondary mirror assembly was designed
and built by R. H. Hildebrand's group at the University of Chicago
and was supported by NSF Grant \# AST 8917950.
\end{acknowledgments}

\clearpage

\begin{figure}
\includegraphics[%
  scale=0.7]{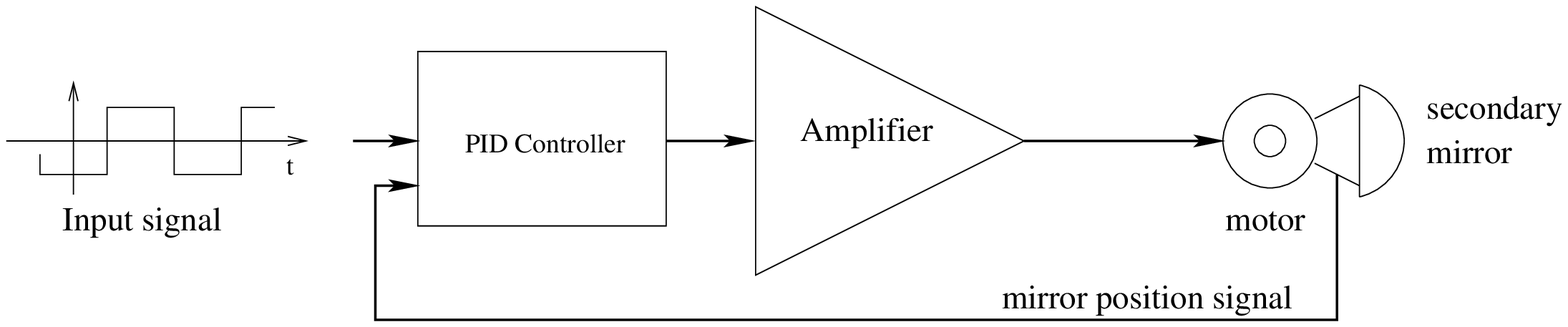}

\caption{\label{cap:pid}The existing chopping secondary mirror assembly at
the CSO. A square wave signal is sent to the input of the PID controller
and compared with the mirror output position signal (from a LVDT).
The resulting processed error signal is sent to a power amplifier
which feeds the positioning motor.}
\end{figure}
\clearpage

\begin{figure}
\includegraphics[%
  scale=0.6]{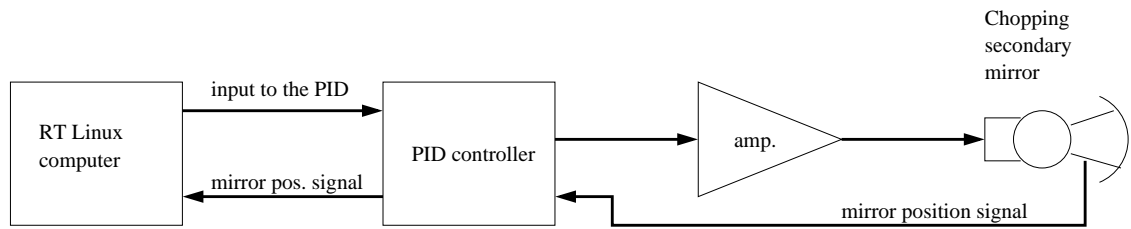}

\caption{\label{cap:new_pid}The new chopping secondary controller. The new
mirror assembly is the same as that of Figure \ref{cap:pid}, but
the input signal is now fed to the PID controller from a RT Linux
computer, which hosts the deconvolution program that determines the
needed input signal.}
\end{figure}

\begin{figure}
\includegraphics[%
  scale=0.5,
  angle=270]{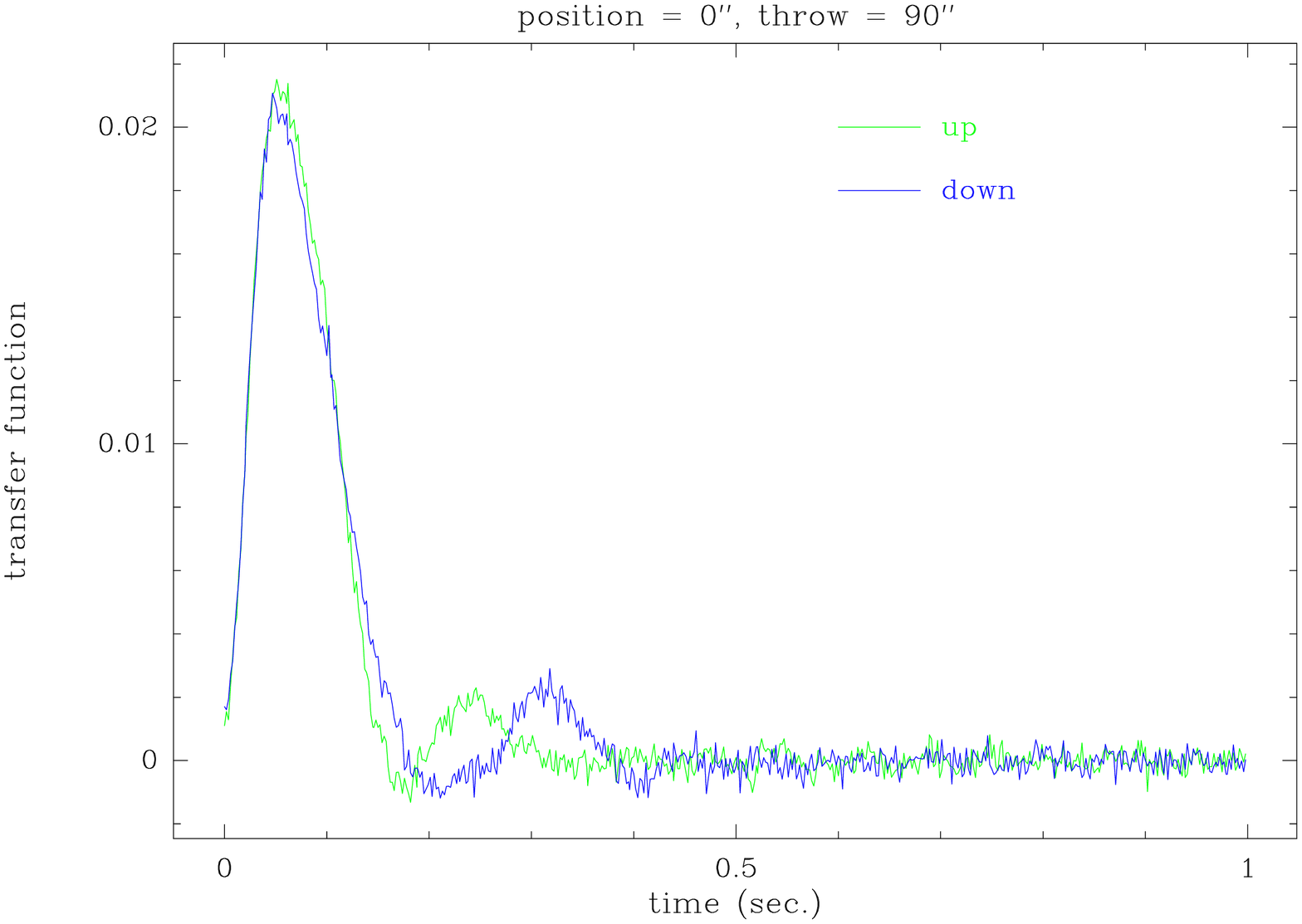}

\vspace{0.2cm}

\includegraphics[%
  scale=0.5,
  angle=270]{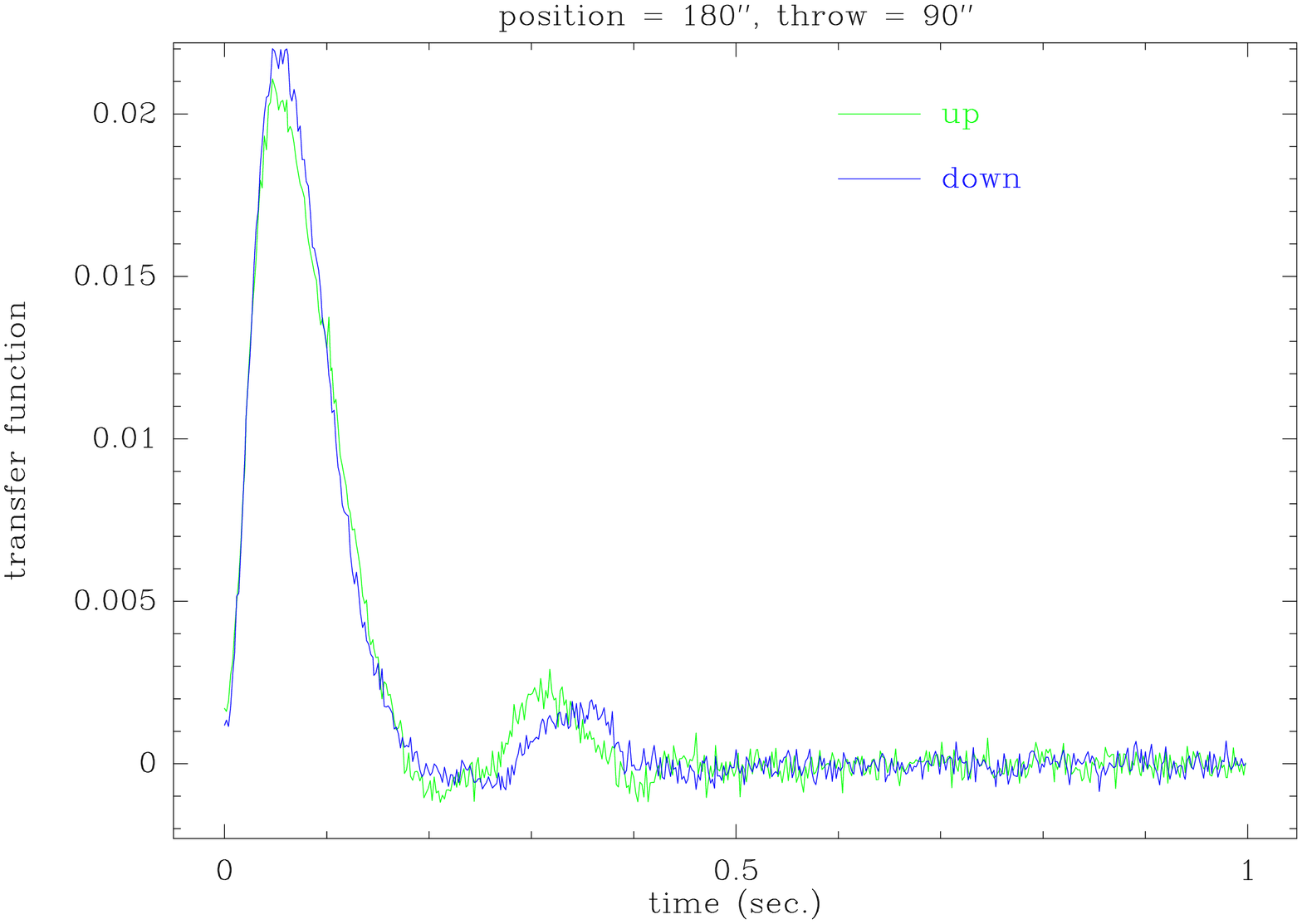}

\caption{\label{cap:tfunc}Effects of the non-linearities as seen through
transfer functions obtained with rising ({}``up'') and falling ({}``down'')
step functions (90 arcseconds in amplitude) at two different rest
positions (0 and 180 arcseconds for the top and bottom graphs, respectively).}
\end{figure}

\begin{figure}
\includegraphics[%
  scale=0.5,
  angle=270]{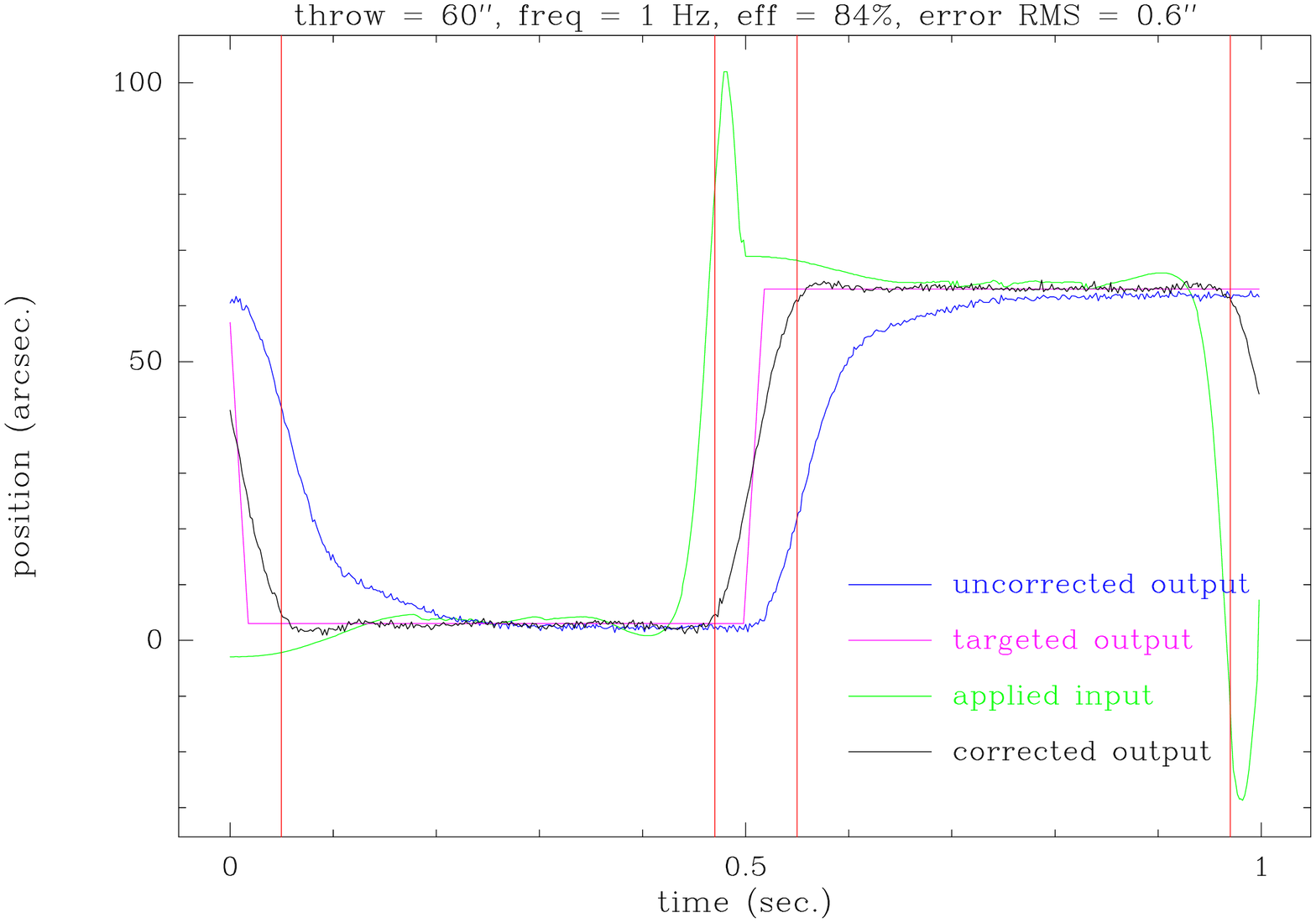}

\vspace{0.2cm}

\includegraphics[%
  scale=0.5,
  angle=270]{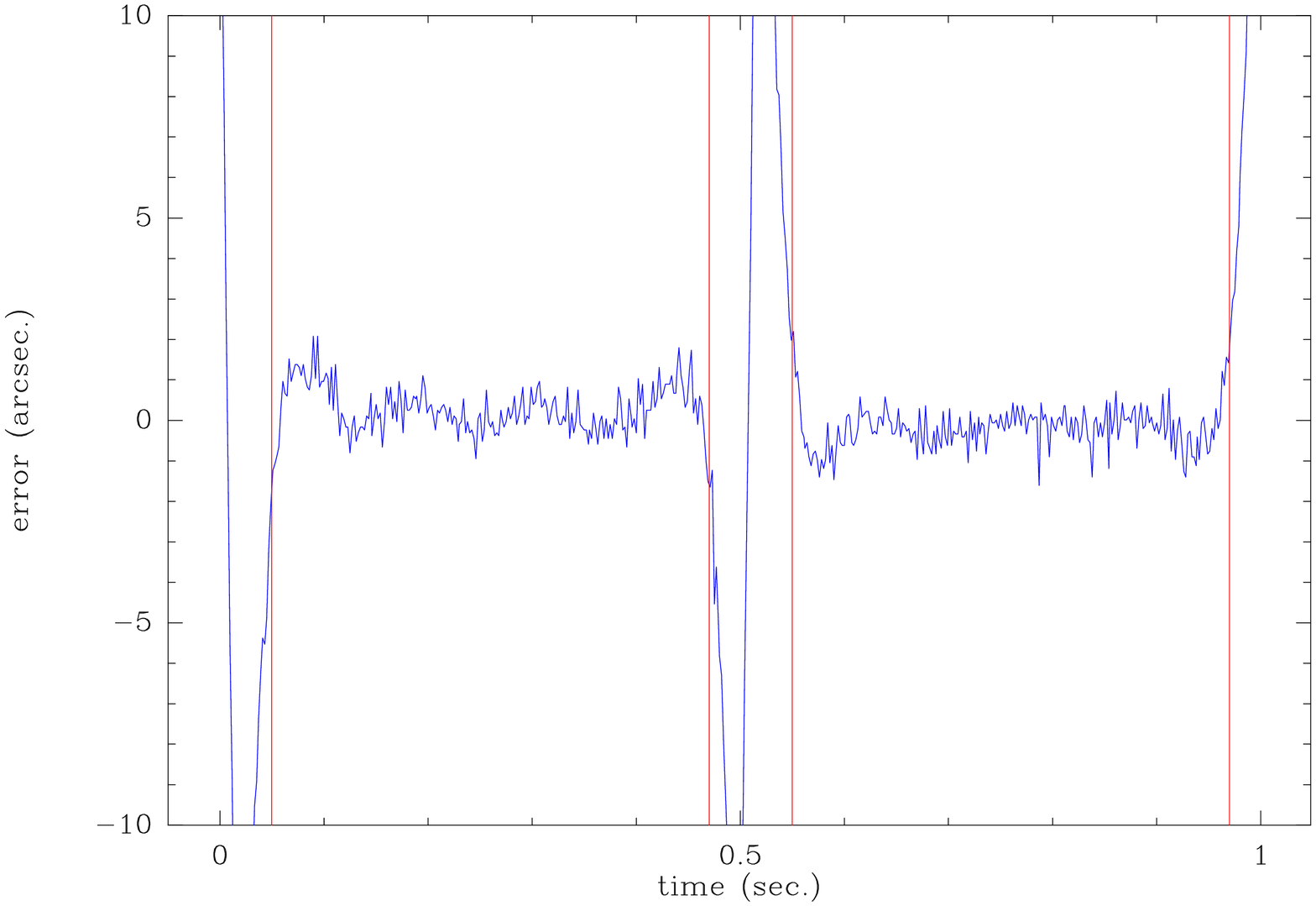}

\caption{\label{cap:60_1hz}Results obtained with our deconvolution technique
for a throw of 60 arcseconds at a frequency of 1 Hz (top). The residual
error signal is plotted in the bottom graph and its RMS value (0.6
arcsecond) was calculated using data points located between the vertical
lines (on the flatter parts of the curve which represent about 84\%
of a period). }
\end{figure}

\begin{figure}
\includegraphics[%
  scale=0.5,
  angle=270]{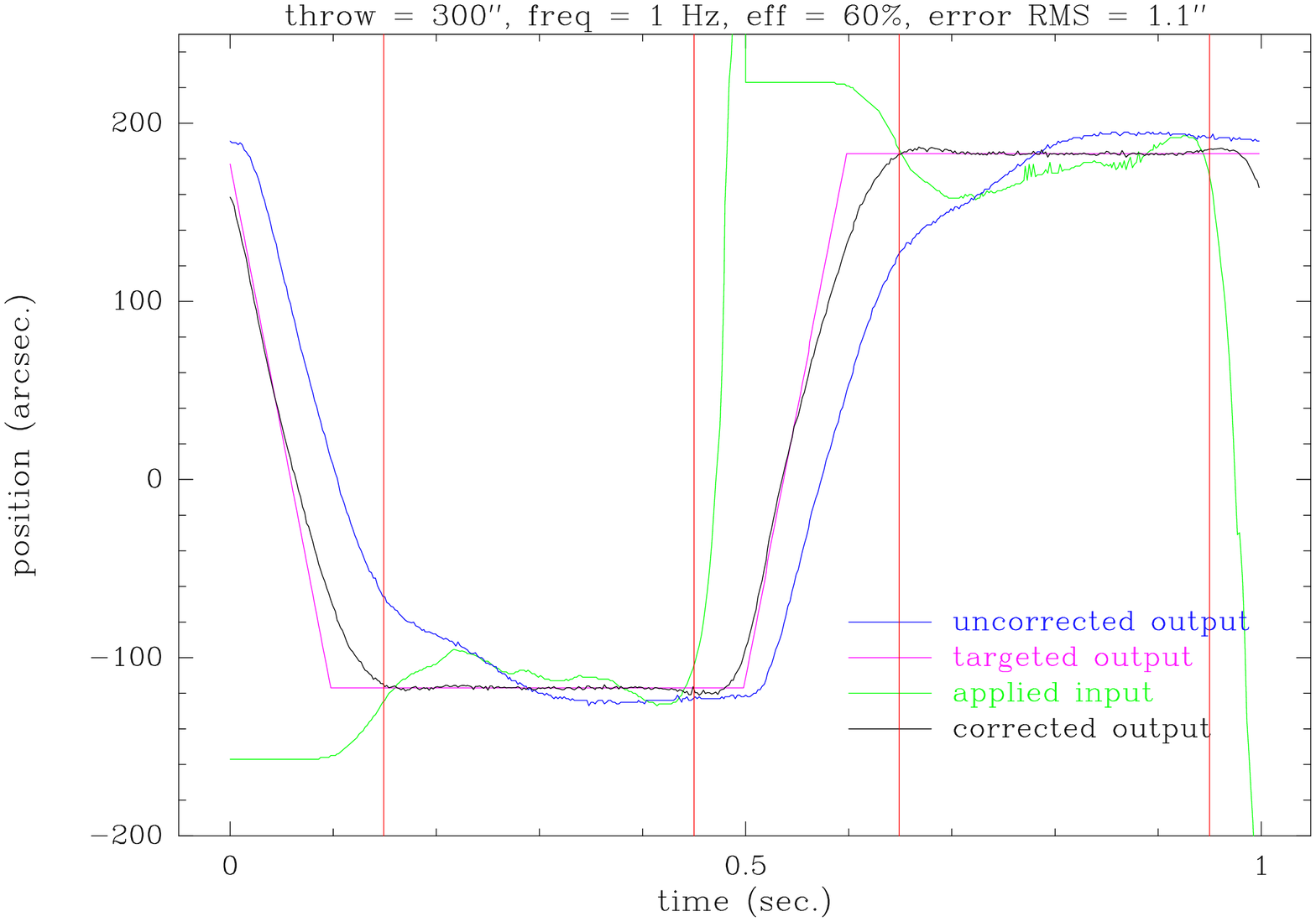}

\vspace{0.2cm}

\includegraphics[%
  scale=0.5,
  angle=270]{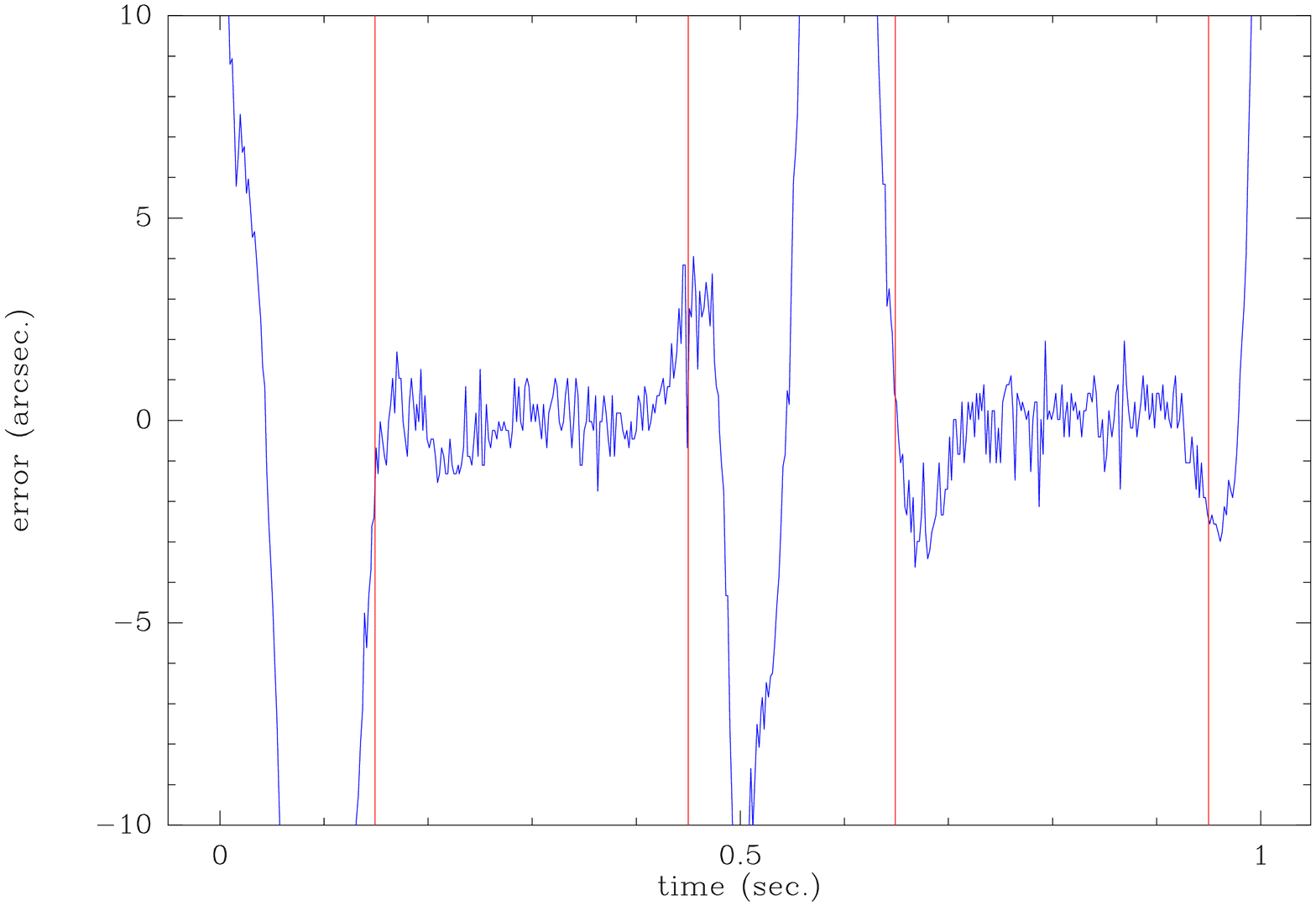}

\caption{\label{cap:300_1hz}Results obtained with our deconvolution technique
for a throw of 300 arcseconds at a frequency of 1 Hz (top). The residual
error signal is plotted in the bottom graph and its RMS value (1.1
arcsecond) was calculated using data points located between the vertical
lines (on the flatter parts of the curve which represent about 60\%
of a period). }
\end{figure}

\begin{figure}
\includegraphics[%
  scale=0.5,
  angle=270]{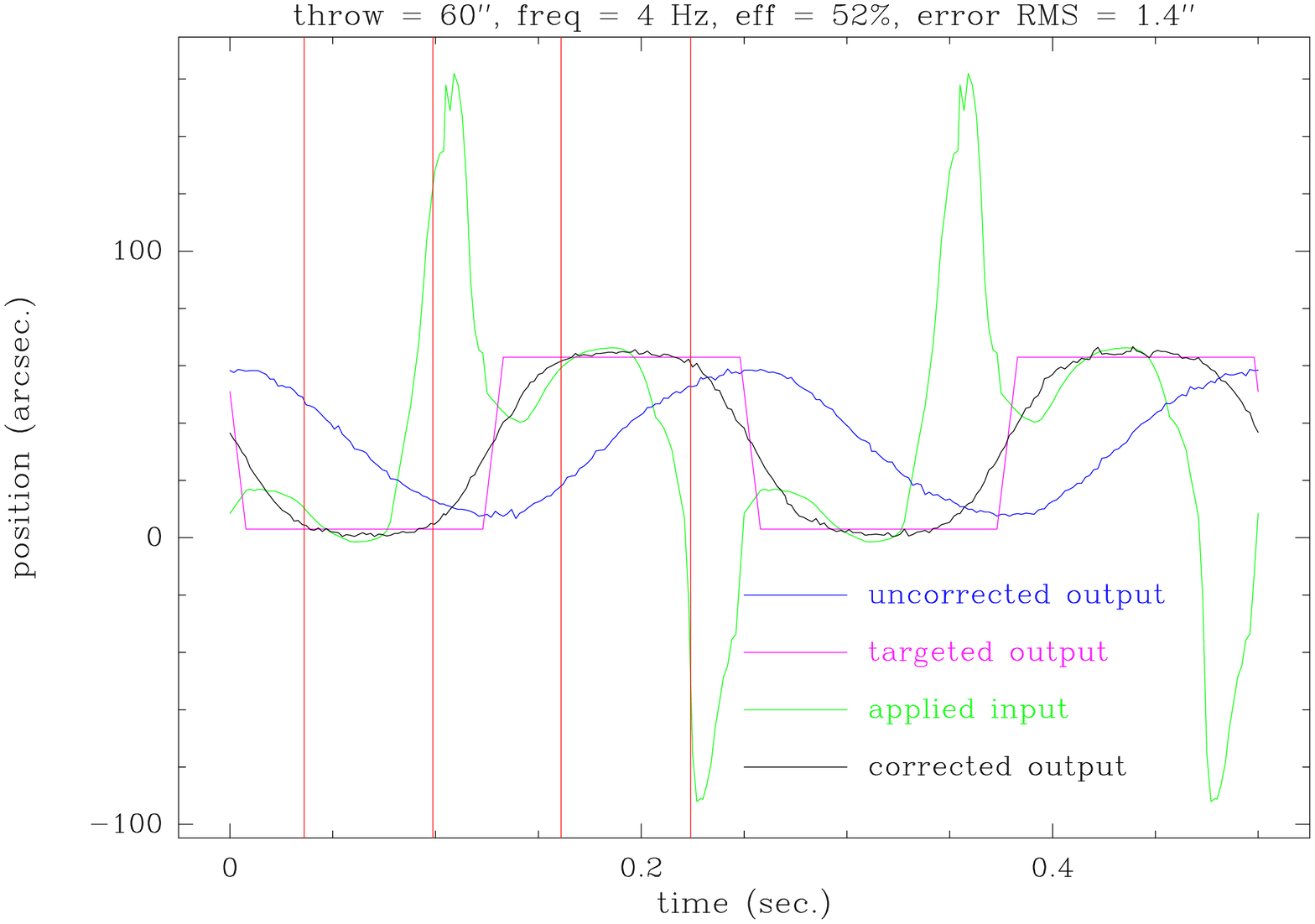}

\vspace{0.2cm}

\includegraphics[%
  scale=0.5,
  angle=270]{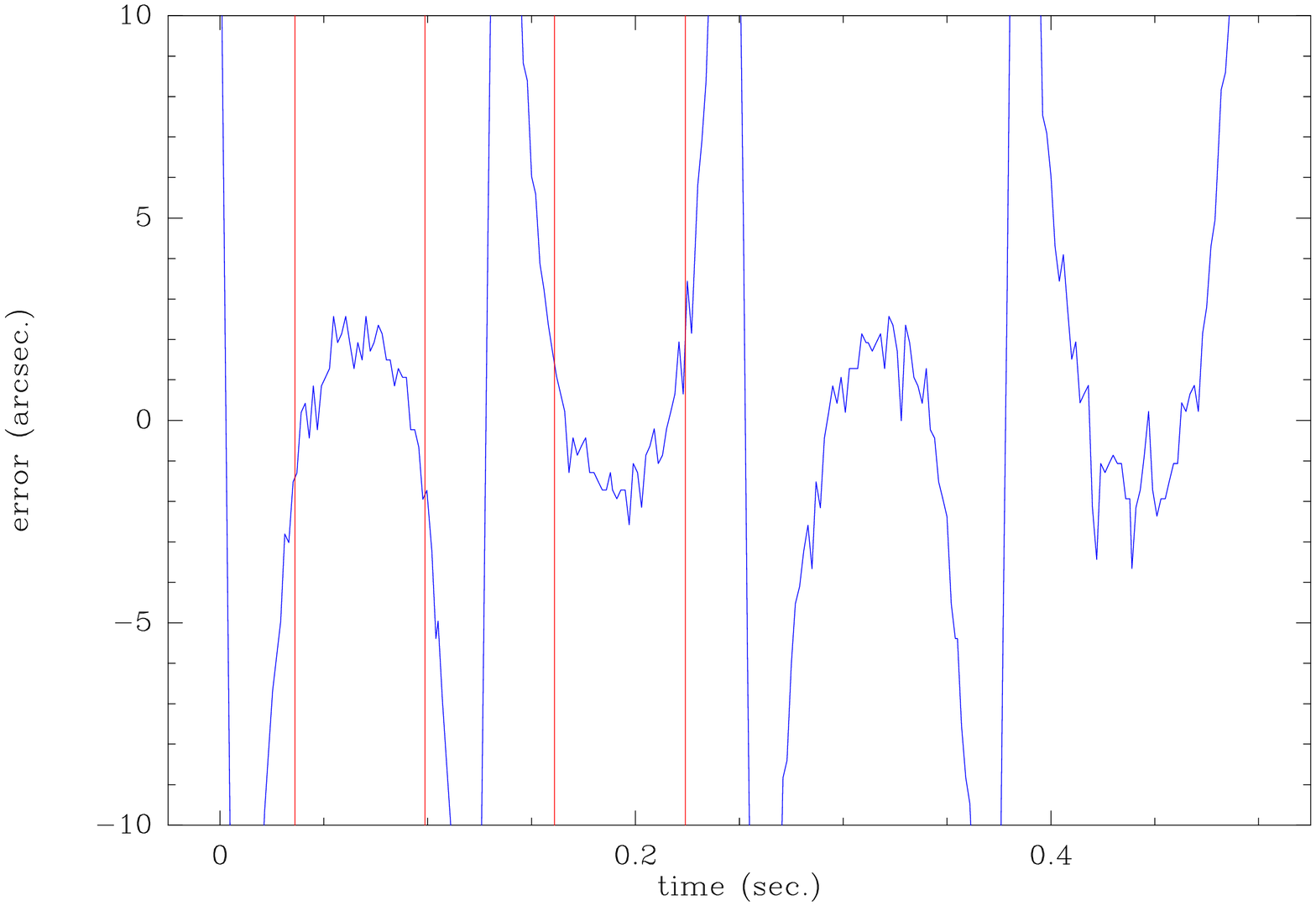}

\caption{\label{cap:60_4hz}Results obtained with our deconvolution technique
for a throw of 60 arcseconds at a frequency of 4 Hz (top). The residual
error signal is plotted in the bottom graph and its RMS value (1.4
arcseconds) was calculated using data points located between the vertical
lines (on the flatter parts of the curve which represent about 52\%
of a period). }
\end{figure}

\clearpage

\section*{Figure Captions}

\noindent Figure 1: The existing chopping secondary mirror assembly
at the CSO. A square wave signal is sent to the input of the PID controller
and compared with the mirror output position signal (from a LVDT).
The resulting processed error signal is sent to a power amplifier
which feeds the positioning motor.\\

\noindent Figure 2: The new chopping secondary controller. The new
mirror assembly is the same as that of Figure \ref{cap:pid}, but
the input signal is now fed to the PID controller from a RT Linux
computer, which hosts the deconvolution program that determines the
needed input signal.\\

\noindent Figure 3: Effects of the non-linearities as seen through
transfer functions obtained with rising ({}``up'') and falling ({}``down'')
step functions (90 arcseconds in amplitude) at two different rest
positions (0 and 180 arcseconds for the top and bottom graphs, respectively).\\

\noindent Figure 4: Results obtained with our deconvolution technique
for a throw of 60 arcseconds at a frequency of 1 Hz (top). The residual
error signal is plotted in the bottom graph and its RMS value (0.6
arcsecond) was calculated using data points located between the vertical
lines (on the flatter parts of the curve which represent about 84\%
of a period). \\

\noindent Figure 5: Results obtained with our deconvolution technique
for a throw of 300 arcseconds at a frequency of 1 Hz (top). The residual
error signal is plotted in the bottom graph and its RMS value (1.1
arcsecond) was calculated using data points located between the vertical
lines (on the flatter parts of the curve which represent about 60\%
of a period). \\

\noindent Figure 6: Results obtained with our deconvolution technique
for a throw of 60 arcseconds at a frequency of 4 Hz (top). The residual
error signal is plotted in the bottom graph and its RMS value (1.4
arcseconds) was calculated using data points located between the vertical
lines (on the flatter parts of the curve which represent about 52\%
of a period). 
\end{document}